\documentclass[5p,times]{elsarticle}%
\usepackage{amssymb}
\usepackage{epsfig}
\usepackage{graphicx}
\usepackage{subfigure}
\usepackage{hyperref}
\usepackage{balance}
\biboptions{numbers,sort&compress}
\journal{Physics Letter A}

\begin{document}
\begin{frontmatter}

\title{Coulomb interaction and semimetal-insulator transition in graphene}
\author{Wei Li and Guo-Zhu Liu}
\address{\small {\it Department of Modern Physics, University of Science and
Technology of China, Hefei, Anhui, 230026, P.R. China}}

\begin{abstract}
The strong Coulomb interaction between massless Dirac fermions can
drive a semimetal-insulator transition in single-layer graphene by
dynamically generating an excitonic fermion gap. There is a critical
interaction strength $\lambda_c$ that separates the semimetal phase
from the insulator phase. We calculate the specific heat and
susceptibility of the system and show that they exhibit distinct
behaviors in the semimetal and insulator phases.
\end{abstract}


\end{frontmatter}


\section{Introduction}

The low-energy properties of graphene have been widely investigated
theoretically and experimentally in recent years \cite{CastroNeto}.
It is well-known that the low-lying elementary excitations of
graphene are massless Dirac fermions, which have linear dispersion
and display quite different behaviors from ordinary electrons with
parabolic dispersion. At half-filling state, the density of states
of massless Dirac fermions vanishes linearly with energy $\omega$
near the Fermi level. Due to this fact, there is essentially no
screening on the Coulomb interaction between Dirac fermions. The
unscreened, long-range Coulomb interaction was argued
\cite{Gonzalez94, Khveshchenko2001, Gusynin02, Herbut06, Son07,
Vafek, Sheehy, Gonzalez96, Sarma2007} to be responsible for a plenty
of unusual physical properties, including the logarithmic velocity
renormalization \cite{Gonzalez94, Son07}, the logarithmic specific
heat correction \cite{Vafek}, the presence of quantum critical point
\cite{Khveshchenko2001, Gusynin02, Son07}, and the marginal Fermi
liquid quasiparticle lifetime \cite{Gonzalez96, Sarma2007}.

When the unscreened Coulomb interaction is sufficiently strong, the
semimetal ground state of graphene may no longer be stable. There
exists an interesting possibility that the massless Dirac
quasiparticles and quasiholes are bound into pairs through the
attractive Coulomb interaction between them. As a consequence, the
massless Dirac fermions acquire a finite mass and the ground state
of graphene becomes insulating. This semimetal-insulator transition
is usually called excitonic instability in the literature
\cite{Khveshchenko2001, Gusynin02, Khveshchenko2006, Hands08,
Drut09, Liu09}. It can be identified as the non-perturbative
phenomenon of dynamical chiral symmetry breaking conventionally
studied in the context of particle physics \cite{Nambu,
Appelquist86, Appelquist88}. Both Dyson-Schwinger equation
\cite{Khveshchenko2001, Gusynin02, Khveshchenko2006, Liu09} and
lattice simulation approaches \cite{Hands08, Drut09} found that such
excitonic instability occurs only when the fermion flavor is less
than a critical value $N_{c}$ and the Coulomb strength is larger
than a critical value $\lambda_{c}$. If we fix the physical fermion
flavor $N = 2$, then the semimetal-insulator transition happens at a
single critical point $\lambda_{c}$.

The effective coupling parameter of Coulomb interaction can be
defined as $\lambda = Ng^{2}/16 = Ne^{2}/(16\epsilon_{0} \hbar
v_{F})$ with $\epsilon_{0}$ being the dielectric constant and
$v_{F}$ being the effective velocity. In the clean limit, the
physical magnitude of this parameter is around 3 or 4 for graphene
in vacuum. In the same limit, we found by solving gap equation to
the leading order of $1/N$ expansion that the critical strength
$\lambda_c \approx 1.85$ \cite{Liu09}. Using analogous gap equation
approach, the critical coupling is found to be $\lambda_{c} \approx
2.08$ and $\lambda_c \approx 4.16$ respectively in Ref.
\cite{Khveshchenko2001} and Ref. \cite{Gusynin02}. In addition, the
Monte Carlo study \cite{Drut09} performed in lattice field theories
found that the critical strength $\lambda_{c} \approx 1.74$ at $N =
2$. Our critical coupling is much more closer in magnitude to that
of Monte Carlo study.

Once a fermion mass gap is generated, the low-energy properties of
graphene fundamentally change. Below the energy scale set by the
fermion gap, the density of states of fermions is substantially
suppressed, which would produce important consequences. It is
interesting to study some observable physical quantities those can
serve as signatures for the existence of excitonic instability. The
effects of dynamical fermion gap have been discussed by several
authors \cite{Gusynin06, Kotov, Asgari2009}. In this paper, we
calculate the specific heat and susceptibility of Dirac fermions and
other low-energy excitations in both semimetal and excitonic
insulator phases. These quantities can be compared with experimental
results and hence may help to understand the physical consequence of
excitonic instability.

In the semimetal phase, the Coulomb interaction is not strong enough
to trigger excitonic pairing instability, but it is strong enough to
produce unusual properties. As found by Vafek \cite{Vafek}, the
long-range Coulomb interaction gives rise to logarithmic
$T$-dependence of fermion specific heat, which is clearly not
behavior of normal Fermi liquid. In this paper, we re-derive the
same qualitative result by a different method. We also calculate the
susceptibility of massless Dirac fermions and show that it also
exhibits logarithmic $T$-dependence due to long-range Coulomb
interaction.

In the insulator phase, the fermion density of states is suppressed
by the excitonic gap. Intuitively, the specific heat and
susceptibility of Dirac fermions should drop significantly from
their corresponding magnitudes in the semimetal phase. Our explicit
computations will show that this is true. However, the massive Dirac
fermions are not the true low-lying elementary excitations in the
insulating state. At the low energy regime, the only degree of
freedom is the massless Goldstone boson which originates from the
dynamical breaking of continuous chiral symmetry. The Goldstone
bosons make dominant contribution to the total specific heat at low
temperature, but make no contribution to the total susceptibility.

In section \ref{sec:model}, we set up the Hamiltonian of the system
and define the physical quantities in which we are interested. We
calculate the free energy, specific heat and susceptibility in
section \ref{sec:clean}. We finally summarize the results and
discuss some relevant problems in section \ref{sec:conclusion}.

\section{Model and Definitions}
\label{sec:model}

The Hamiltonian of massless Dirac fermions in single layer graphene
is given by
\begin{eqnarray}\label{eq:hamiltonian}
H_{0} &=& v_{F}\sum_{\sigma=1}^{N}\int_{\mathbf{r}}
\bar{\psi}_{\sigma}(t,\mathbf{r})i\mathbf{\gamma}\cdot\mathbf{\nabla}
\psi_{\sigma}(t,\mathbf{r}),  \\
H_{\mathrm{C}} &=& \frac{1}{4\pi}\sum_{\sigma,\sigma^{\prime}}^{N}
\int_{\mathbf{r},\mathbf{r}^{\prime}}
\bar{\psi}_{\sigma}(t,\mathbf{r})\gamma_{0}\psi_{\sigma}(t,\mathbf{r})
U_{C}(t,t^{\prime},\mathbf{r},\mathbf{r}^{\prime})
\nonumber \\&&
\times\,\bar{\psi}_{\sigma^{\prime}}(t^{\prime},\mathbf{r}^{\prime})\gamma_{0}
\psi_{\sigma^{\prime}}(t^{\prime},\mathbf{r}^{\prime}),
\end{eqnarray}
where the Coulomb interaction potential \cite{Gusynin02} is
\begin{eqnarray}\label{eq:V}
U_{C}(t,\mathbf{r}) =
g^{2}\int\frac{d\omega}{2\pi}\frac{d^2\mathbf{k}}{2\pi}
\frac{e^{-i\omega t + i\bf{k}\cdot\bf{r}}}{|{\bf
k}|+\Pi(\omega,\bf{k})},
\end{eqnarray}
where $g^{2} = e^{2}/\epsilon_{0}\hbar v_{F}$. As mentioned in
Introduction, it is convenient to define a dimensionless Coulomb
coupling as $\lambda = Ng^{2}/16$. Usually, Dirac fermion in two
spatial dimensions is described by two-component spinor field whose
$2\times 2$ representation can be formulated by Pauli matrices
$\gamma_{\mu}=(\sigma_2,i\sigma_3,i\sigma_1)$. However, it is not
possible to define a $2\times 2$ matrix that anticommutes with all
these matrices. Therefore, there is no chiral symmetry in this
representation. Here, we adopt four-component spinor field $\psi$ to
describe the massless Dirac fermion \cite{Appelquist86, Drut09}. The
conjugate spinor field is defined as $\bar{\psi} =
\psi^{\dagger}\gamma_{0}$. The $4\times 4$ $\gamma$-matrices can be
defined as
$\gamma_{\mu}=(\sigma_3,i\sigma_1,i\sigma_2)\otimes\sigma_{3}$,
which satisfy the standard Clifford algebra $\{\gamma_{\mu},
\gamma_{\nu}\} = 2g_{\mu\nu}$ with metric $g_{\mu\nu} =
\mathrm{diag}(1,-1,-1)$.  Obviously, there are two $4\times 4$
matrices
\begin{eqnarray}
  \gamma_{3} = i\left(
  \begin{array}{cc}
    0 & I \\
    I & 0 \\
  \end{array}
\right)
,\,\,\,\,\,\,\
    \gamma_{5} = i\left(
  \begin{array}{cc}
    0 & I \\
    -I & 0 \\
  \end{array}
\right),\nonumber
\end{eqnarray}
which anticommute with all $\gamma_{\mu}$. The total Hamiltonian
preserves a continuous U(2N) chiral symmetry $\psi \rightarrow
e^{i\alpha\gamma_{3,5}}\psi$. The mass term  generated by excitonic
pairing will break this global chiral symmetry dynamically to
subgroup $U(N)\times U(N)$. Meanwhile, according to the Goldstone
theorem, there appear massless Goldstone bosons due to the breaking
of continuous chiral symmetry. These bosons are the only gapless
excitations in the symmetry broken phase and hence play an important
role in determining the low-energy behaviors of the system. Although
the physical fermion flavor is actually $N = 2$, in the following we
consider a general $N$ in order to perform $1/N$ expansion. For
convenience, we work in units where $\hbar = k_{B} = v_{F} = 1$
throughout the paper.

The electronic structure of graphene is very special in that the
$\pi$-conduction bands and $\pi^{*}$-valence bands touch at two
inequivalent $K$ points. This is the reason why the low-energy
fermionic excitations have a linear dispersion. When the strong,
long-range Coulomb interaction opens an excitonic gap at the Dirac
point, the chiral symmetry of total Hamiltonian is broken,
resembling the non-perturbative phenomenon of dynamical chiral
symmetry breaking in QED$_{3}$ \cite{Appelquist88}. This mechanism
was first proposed in graphene by Khveshchenko
\cite{Khveshchenko2001} and has been extensively studied
\cite{Gusynin02, Khveshchenko2006, Hands08, Drut09, Liu09} in the
following years.

Gusynin \emph{et} \emph{al.} discussed the influence of excitonic
fermion gap on various transport quantities, including electrical
and Hall conductivity \cite{Gusynin02, Gusynin06}. The results were
compared directly with the experiments in graphene. Recently, Kotov
\emph{et} \emph{al.} studied the effect of fermion gap on the
interacting potential \cite{Kotov} and found an effective weak
confinement of fermions. They also argued that the massive phase
exhibits much more interesting behavior than the massless one. The
effect of fermion gap on quasiparticle lifetime and spectral
function was discussed in \cite{Asgari2009}. This kind of excitonic
instability may also exist in other correlated electron systems than
graphene. For instance, it was suggested by one of the authors that
such instability can provide a qualitative understanding on the
field-induced thermal metal-insulator transition observed in the
vortex state of high temperature cuprate superconductor
\cite{Liu09HTC}.

In this paper, we calculate the specific heat and susceptibility by
including the effect of Coulomb interaction in both the semimetal
and insulator phases. These are physical quantities those can be
measured by experiments and hence can help us to build interesting
connections between theoretical predictions and experimental facts.
Technically, we will follow the procedures utilized in the paper of
Kaul and Sachdev \cite{Kaul}. In this framework, all propagators and
correlation functions are written in the Matsubara imaginary time
formalism.

At finite temperature, the fermion propagator is
\begin{equation}\label{eq:propagator}
G(i\omega_{n},\mathbf{k}) =
\frac{1}{i\omega_{n}\gamma_{0} - \mathbf{\gamma}\cdot\mathbf{k} - m},
\end{equation}
where $\omega_{n} = (2n+1)\pi T$ is the fermion frequency. Although
generally the excitonic fermion gap should depend on momentum,
energy, and temperature, we assume a constant mass gap $m$
throughout the paper to simplify calculations.

The bare Coulomb interaction function is simply
\begin{equation}
D_{0}(\omega_{m},\mathbf{q}) = \frac{g^{2}}{2\mathbf{q}} =
\frac{\lambda}{\frac{N}{8} \mathbf{q}}.
\end{equation}
In an interacting electron gas, the collective excitations screen
the bare Coulomb interaction and convert $D_{0}$ to
\begin{equation}
D(\omega_{m},\mathbf{q},T) =
\frac{1}{\frac{N}{8}}\frac{1}{\frac{\mathbf{q}}{\lambda} +
\frac{8}{N}\Pi(\omega_{m},\mathbf{q},T)},
\end{equation}
where the polarization function $\Pi$ is defined as
\begin{equation}\label{eq:pi}
\Pi(\omega_{m},\mathbf{q},T) = -N T
\sum_{\omega_{n}}\int\!\frac{d^{2}k}{(2\pi)^{2}}
\frac{\mathrm{Tr}[\gamma_{0} k\!\!\!/\gamma_{0}(q\!\!\!/+k\!\!\!/)]}
{k^2(q+k)^{2}},
\end{equation}
with $q_{0} \equiv \omega_{m} = 2m\pi T$ and $k_{0}
\equiv \omega_{n} = (2n+1)\pi T$.

The first two orders in $1/N$ expansion of free energy $\mathcal{F}$
are
\begin{equation}\label{eq:freeeE}
\mathcal{F} = Nf^{0f}+f^{1f},
\end{equation}
where $f^{0f}$ is the leading, noninteracting term and $f^{1f}$ the
corrections from the Coulomb interaction. The leading term of
fermion free energy is defined as $f^{0f} =
T\sum\limits_{\omega_n}\int\frac{d^2\mathbf{k}}{(2\pi)^2}
\ln[G(i\omega_{n},\mathbf{k})]$. To calculate the free energy, we
will first sum over the Matsubara frequencies $\omega_{n}$ and then
perform the integration over the intermediate variables and momentum
$k$, dropping all terms those are independent of temperature and
volume \cite{Kapusta}. The volume factor is neglected throughout
this paper and we only consider free energy in unit volume. The
interaction correction to free energy is given by
\begin{equation}
\mathcal{F}(T)=
T\sum_{\omega_m}\int\frac{d^2q}{(2\pi)^2}\ln[D^{-1}].
\end{equation}
We choose the zero temperature free energy $\mathcal{F}(T = 0)$ as
the reference free energy \cite{Vafek, Asgari}, and then define the
following regularized free energy
\begin{eqnarray}
\label{eq:delta_f}
f^{1f} &\equiv& \mathcal{F}(T)-\mathcal{F}(T=0)
\nonumber \\ &=&
T\sum_{\omega_{m}}\int\frac{d^2\mathbf{q}}{(2\pi)^2}
\ln\left[\frac{D^{-1}(\omega_{m},\mathbf{q},T)}
{D^{-1}(\omega_{m},\mathbf{q},T=0)}\right].
\end{eqnarray}
Here, we follow the strategy of Ref. \cite{Kaul} and introduce a
magnetic field $H$. For fermions, the field shifts frequency as
$\omega_{n} \rightarrow \omega_{n}-\theta H$, where $\theta = \pm
1$. The specific heat $C_{V}$ and susceptibility $\chi_{f}$ can be
defined as
\begin{eqnarray}
\label{eq:tq}
C_{V} = -T \frac{\partial^{2}\mathcal{F}}{\partial T^{2}}
=N C_{V}^{0f} + C_{V}^{1f},\\
\chi_{f} = \frac{\partial^{2}\mathcal{F}}{\partial H^{2}}\Big|_{H=0}
=N \chi_{f}^{0f} + \chi_{f}^{1f},
\end{eqnarray}
which are divided to free and interaction terms, respectively. For a
normal Fermi liquid, the specific heat and susceptibility should
behave as $C_{V}\propto T^{2}$ and $\chi_{f}\propto T$ according to
the analysis in Ref.~\cite{Kaul, Chubukovprb2005}. If we write the
specific heat as $C_{V} = \mathcal{A}_{C_V} T^{2}$, then
$\mathcal{A}_{C_V}$ should be
\begin{equation}
\label{eq:acv}
\mathcal{A}_{C_V}=N\mathcal{A}^{0f}_{C_V}+\mathcal{A}^{1f}_{C_V}.
\end{equation}
Similarly, the susceptibility can also be written as $\chi_{f} =
\mathcal{A}_{\chi_{f}} T$ with $\mathcal{A}_{\chi_{f}}$ being
\begin{equation}
\label{eq:achi}
\mathcal{A}_{\chi_{f}}=N\mathcal{A}^{0f}_{\chi_{f}}+\mathcal{A}^{1f}_{\chi_{f}}.
\end{equation}
In the presence of fermion mass $m$, the specific heat
(susceptibility) no longer behaves as $\propto T^{2}$ ($\propto T$).
However, in order to make direct comparison, we still express
specific heat (susceptibility) in terms of $\mathcal{A}_{C_V}$
($\mathcal{A}_{\chi_{f}}$), which will depend on temperature $T$.
The definitions presented in this section will be used to calculate
the free energy, specific heat, and susceptibility in the next
section.

\section{Specific heat and susceptibility}
\label{sec:clean}

\subsection{Leading terms}

\begin{figure}[ht]
  \centering
    \subfigure[]{
    \label{fig:spheat_0} 
    \includegraphics[width=2.7in]{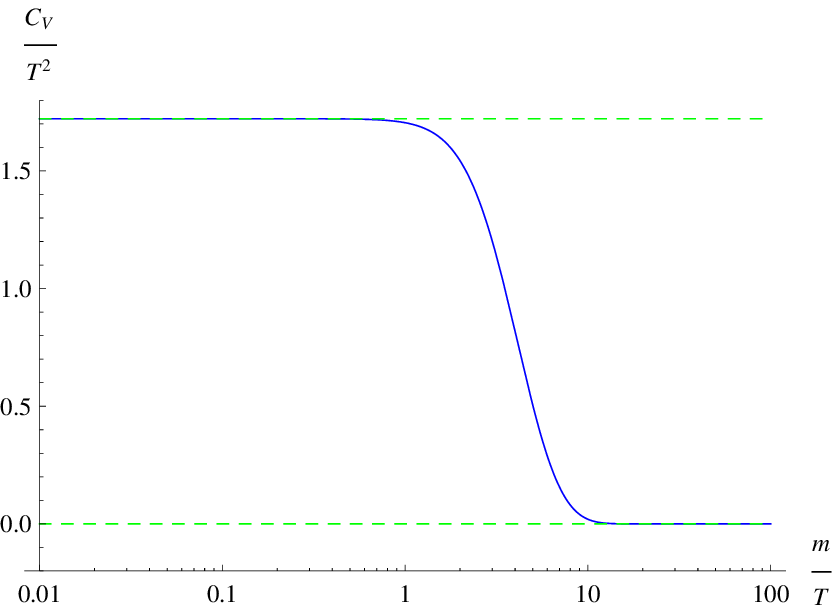}}
    \subfigure[]{
    \label{fig:sus_0} 
    \includegraphics[width=2.7in]{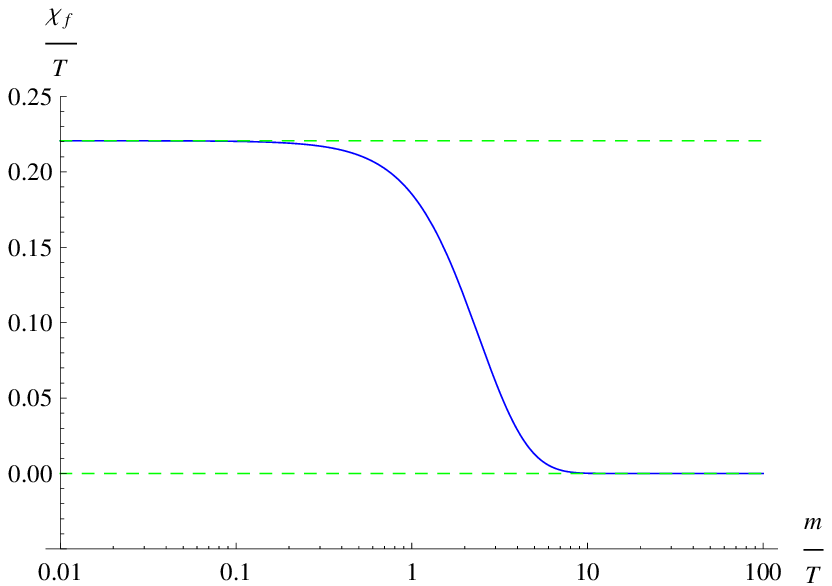}}
\caption{(a) The dependence of fermion specific heat on fermion gap
$\frac{m}{T}$. The top dashed line corresponds to the
non-interacting term $\mathcal{A}^{0f}_{C_V} =
\frac{9\zeta(3)}{2\pi}$ at $m=0$. (b) The dependence of fermion
susceptibility on fermion gap $\frac{m}{T}$. The top dashed line
corresponds to the non-interacting term $\mathcal{A}^{0f}_{\chi_f} =
\frac{\ln 2}{\pi}$ at $m=0$. It is obvious that fermion gap
suppresses fermion specific heat and susceptibility strongly.}
 \label{fig:free_0} 
\end{figure}

In the presence of a constant fermion mass $m$, the noninteracting
free energy is
\begin{eqnarray}\label{eq:free_m0}
f^{0f}(m) &=&
T\sum\limits_{\omega_n}\int\frac{d^2\mathbf{k}}{(2\pi)^2}
\ln[G(i\omega_{n},\mathbf{k})] \nonumber \\
&=& -T\int \frac{d^{2}\mathbf{k}}{2\pi^{2}} \ln \left[1 +
e^{-\frac{\sqrt{\mathbf{k}^{2}+m^{2}}}{T}\pm
i\theta\frac{H}{T}}\right] \nonumber \\&=& -
\frac{1}{\pi}\Bigg\{\frac{1}{3}\left(\frac{m}{T}\right)^{3} +
\frac{1}{4}\frac{m}{T}\mathrm{Li}_{2}\left[-e^{\frac{m}{T}\pm
i\theta\frac{H}{T}}\right] \nonumber \\&& -
\frac{1}{4}\mathrm{Li}_{3}\left[-e^{\frac{m}{T}\pm
i\theta\frac{H}{T}}\right]\Bigg\}T^{3}.
\end{eqnarray}
Here, $\mathrm{Li}_{2}$ and $\mathrm{Li}_3$ are polylogarithmic
functions. It is easy to get the following noninteracting term for
fermion specific heat
\begin{eqnarray}\label{eq:spheat_0}
\mathcal{A}^{0f}_{C_V} &=&
-\frac{1}{\pi}\Bigg\{\frac{\left(\frac{m}{T}\right)^{3}}{1+e^{-\frac{m}{T}}}
-3\left(\frac{m}{T}\right)^{2} \ln\left[1+e^{\frac{m}{T}}\right]
\nonumber \\&&
-6\left(\frac{m}{T}\right)\mathrm{Li}_{2}\left[-e^{\frac{m}{T}}\right]
+ 6 \mathrm{Li}_{3}\left[-e^{\frac{m}{T}}\right]\Bigg\}.
\end{eqnarray}
This function is plotted in Fig. \ref{fig:spheat_0}. Taking the $m=0$
limit of Eq. (\ref{eq:spheat_0}), the specific heat in the
semimetal phase is
\begin{equation}
C_{V}^{0f} = \frac{9\zeta(3)}{2\pi}T^{2},
\end{equation}
with $\mathcal{A}^{0f}_{C_V} = \frac{9\zeta(3)}{2\pi}$.
Similarly, the noninteracting susceptibility is given as
\begin{equation}\label{eq:sus_0}
\mathcal{A}^{0f}_{\chi_{f}} =
\frac{1}{\pi}\Big\{\left(\frac{m}{T}\right)\frac{1}{1+e^{-\frac{m}{T}}}
+ \ln\left[1+e^{\frac{m}{T}}\right]\Big\},
\end{equation}
which is plotted in Fig. \ref{fig:sus_0}. Taking the $m=0$
limit of Eq. (\ref{eq:sus_0}), the susceptibility in the
semimetal phase is
\begin{equation}
\chi_{f}^{0f} = \frac{\ln 2}{\pi}T,
\end{equation}
with $\mathcal{A}^{0f}_{\chi_{f}} = \frac{\ln 2}{\pi}$.

From Eq. (\ref{eq:spheat_0}), Eq. (\ref{eq:sus_0}), and Fig.
\ref{fig:free_0}, it is easy to see that the $T$-dependence of
specific heat and susceptibility in the insulating phase differs
significantly from the corresponding $\propto T^{2}$ and $\propto T$
behaviors in the semimetal phase. This is not unexpected because the
excitonic gap strongly suppresses the fermionic excitations at low
temperature.

However, although the low-energy fermion excitations are strongly
suppressed in the insulating phase, there exists another kind of
gapless excitation: Goldstone boson. The presence of gapless
Goldstone bosons is the characteristic property of excitonic
instability. They are composed of Dirac fermions (quasiparticles)
and anti-fermions (quasiholes), but carry no electric charge
themselves. The Goldstone bosons do not contribute to the
susceptibility because they do not couple to external magnetic field
$H$, but they do contribute to the total specific heat of the
system. In particular, the free energy of Goldstone bosons is
\begin{equation}\label{eq:fe_g}
f^{G} = T\sum_{\omega_n}\int\frac{d^{2}k}{4\pi^{2}}[\ln(k^{2} +
\omega_{n}^{2})] = -\frac{2\zeta(3)}{\pi}T^3,
\end{equation}
while the corresponding specific heat is
\begin{equation}
C_{V}^{G} = 12\frac{\zeta(3)}{\pi}T^{2},
\end{equation}
which has been obtained in Ref. \cite{Liuconfine}.

Apparently, a $\propto T^{2}$ term of specific heat appears in both
the semimetal phase and the insulating phase. It seems that these
two phases have similar specific heat, albeit contributed from
different elementary excitations. However, such similarity actually
does not exist because it disappears once the interaction correction
to free energy is incorporated.

\subsection{Interaction corrections}

We now include the interaction correction to the free energy. Note
the Goldstone bosons are neutral, so the Coulomb interaction only
affects the free energy of Dirac fermions. To calculate the free
energy $f^{1f}$, we should first know the polarization function. In
the presence of finite fermion mass $m$ and external magnetic field
$H$, the polarization function $\Pi(\omega_{m},\mathbf{q},T)$ can be
calculated by the methods presented in \cite{Liu09, Dorey,
Liuconfine}. Here we only write down the final expression:
\begin{eqnarray}\label{eq:pifull}
&&\Pi(\omega_{m},\mathbf{q},T)
\nonumber \\&& =
\frac{NT}{\pi}\int_{0}^{1}dx\ln[4D_{m}]
\nonumber \\&&\,\, -\frac{N}{4\pi}\int_{0}^{1}dx \frac{(1-2x)\omega_{m}
\sin(\frac{1}{T}x \omega_{m} + \frac{1}{T}\theta H)}{D_{m}}
\nonumber \\&&\,\, -\frac{N}{2\pi}\int_{0}^{1}dx
\frac{m^{2}+x(1-x)\omega_{m}^{2}}{E_{f}}
\frac{\sinh(\frac{1}{T}E_f)}{D_{m}},
\end{eqnarray}
at finite temperature. Here we introduced the following abbreviated
notations
\begin{eqnarray}
E_{f} &=& \sqrt{m^{2}+x(1-x)(\mathbf{q}^{2} + \omega_{m}^{2})}, \\
D_{m} &=& \cosh^{2}(\frac{ E_{f}}{2T}) - \sin^{2}(\frac{x \omega_{m}
+ \theta H}{2T}).
\end{eqnarray}
The zero-temperature limit of polarization function (Eq.
(\ref{eq:pifull})) is
\begin{equation}\label{eq:pi_0}
\Pi(\omega_{m},\mathbf{q},T=0) = \frac{N}{\pi}\int_{0}^{1}
dx\frac{x(1-x)\mathbf{q}^{2}}{E_{f}}.
\end{equation}
Using Eq. (\ref{eq:pifull}) and Eq. (\ref{eq:pi_0}), the free energy
$f^{1f}$ (Eq. (\ref{eq:delta_f})) can be directly computed.

In the semimetal phase with $m=0$, the polarization function is
\begin{eqnarray}\label{eq:pi_t0}
&&\Pi(\omega_{m},\mathbf{q},T,m=0) =
\frac{\mathbf{q}^{2}}{\mathbf{q}^{2} +\omega_{m}^{2}}\frac{N T}{\pi}
\int_{0}^{1}dx \nonumber \\&&\,\,
\times\ln\left[2\left(\cosh(\frac{E_{f0}}{T}) +
\cos(\frac{x\omega_{m}}{T} + \theta\frac{H}{T})\right)\right],
\end{eqnarray}
at finite temperature with $E_{f0} =\sqrt{x(1-x)(\mathbf{q}^{2}
+\omega_{m}^{2})}$ and
\begin{equation}\label{eq:pi_00}
\Pi(\omega_{m},\mathbf{q},T=0,m=0) =
\frac{N}{8}\frac{\mathbf{q}^{2}}{\sqrt{\mathbf{q}^{2}
+\omega_{m}^{2}}},
\end{equation}
at zero temperature. Using these expressions, the free energy of
Dirac fermion is written as
\begin{equation}\label{eq:delta_f0}
f^{1f} =
\frac{4T^3}{\pi^3}\int_{\delta}^{\frac{\Lambda}{T}}\!\mathbf{q}\,d\mathbf{q}
\int_{0}^{\infty}dy \int_{0}^{1}dx
\{K(x,y,\mathbf{q})+J(x,y,\mathbf{q})\},\nonumber
\end{equation}
where
\begin{eqnarray}
K(x,y,\mathbf{q}) &=&
\frac{\exp\left[-2\mathbf{q}\sqrt{x(1-x)(1+y^{2})}\right]}
{\pi(1+y^{2})\left[\frac{1}{\lambda}+\frac{1}{\sqrt{1+y^{2}}}\right]}
,  \\
J(x,y,\mathbf{q}) &=&
\frac{2\exp\left[-\mathbf{q}\sqrt{x(1-x)(1+y^{2})}\right]}
{\pi(1+y^{2})\left[\frac{1}{\lambda}+\frac{1}{\sqrt{1+y^{2}}}\right]}
\nonumber \\&& \times\cos(xy\mathbf{q} + \theta\frac{H}{T}).
\end{eqnarray}
Here, a variable $y \equiv \frac{\omega}{\mathbf{q}}$ is introduced,
with $\omega$ being the continuous form of $\omega_{m}$ when $T
\rightarrow 0$. For finite $y$, $K(x,y,\mathbf{q})$ damps rapidly
with growing $y$, so $y \sim 0$ makes the dominant contribution to
the free energy. We can expand the function $K(x,y,\mathbf{q})$ near
this point and obtain
\begin{eqnarray}
\label{eq:deltaf1} f^{1f}_1 &=&
\frac{4T^3}{\pi^3}\int_{\delta}^{\frac{\Lambda}{T}}\mathbf{q}\,d\mathbf{q}
\int_{0}^{\infty}dy \int_{0}^{1}dx\,K(x,y,\mathbf{q})
\nonumber\\&=&
\frac{8T^3}{\pi^3}\int_{\delta}^{\frac{\Lambda}{T}}\mathbf{q}\,d\mathbf{q}
\int_{0}^{\infty}dy \int_{0}^{\frac{1}{2}}dx
\nonumber\\&&\times
\frac{\exp\left[-\mathbf{q}\sqrt{(1-4x^2)(1+y^{2})}\right]}{(1+y^{2})
\left[\frac{1}{\lambda}+\frac{1}{\sqrt{1+y^{2}}}\right]}
\nonumber\\&\approx&
\frac{8T^3}{\pi^3}\int_{\delta}^{\frac{\Lambda}{T}}\mathbf{q}\,d\mathbf{q}
\int_{0}^{\infty}dy \frac{1}{(1+y^{2})\left[\frac{1}{\lambda} +
\frac{1}{\sqrt{1+y^{2}}}\right]}
\nonumber \\&&\times
\int_{0}^{\frac{1}{2}}\frac{dx}{\exp\left[2\mathbf{q}\sqrt{x}\sqrt{(1+y^{2})}\right]}
\nonumber \\&=&
\frac{4}{\pi^3}\eta(\lambda)\,T^{3}\ln\frac{\Lambda}{T},
\end{eqnarray}
where
\begin{equation}
\eta(\lambda) = 1 +
\frac{\tan^{-1}\left[\frac{\sqrt{1-\lambda^2}}{\lambda}\right]}{\lambda
\sqrt{1-\lambda^2}}-\frac{\pi}{2\lambda }.
\end{equation}
Comparing with $K(x,y,\mathbf{q})$, the form of $J(x,y,\mathbf{q})$
is more complicated owing to the cosine term $\cos(xy\mathbf{q} +
\theta H/T)$. The computation becomes difficult if we make Taylor
expansion of the cosine function. By plotting the dependence of
function $J(x,y,\mathbf{q})$ on its variables, we found that the
dominant regime is $x \sim 0,\, y\sim 0$. Hence, we simply take
$J(x,y,\mathbf{q})$ as
\begin{equation}
J(x,y,\mathbf{q}) \approx \frac{2\exp\left[-\mathbf{q}\sqrt{x(1
+y^{2})}\right]}
{(1+y^{2})\left[\frac{1}{\lambda}+\frac{1}{\sqrt{1+y^{2}}}\right]}\cos(\theta
\frac{H}{T}),
\end{equation}
which then leads to
\begin{eqnarray}\label{eq:deltaf2}
f^{1f}_{2} &=& \frac{4T^3}{\pi^3}
\int_{\delta}^{\frac{\Lambda}{T}}\mathbf{q}\,d\mathbf{q}\int_{0}^{\infty}dy
\int_{0}^{1}dx\,J(x,y,\mathbf{q})
\nonumber \\
&\approx& \frac{16}{\pi^3}\eta(\lambda)\cos(\theta
\frac{H}{T})\,T^{3}\ln\frac{\Lambda}{T}.
\end{eqnarray}
Taking $H = 0$, the total free energy now has the form
\begin{equation}
\label{eq:free_ln} f^{1f}=
\frac{20}{\pi^3}\eta(\lambda)\,T^{3}\ln\frac{\Lambda}{T}.
\end{equation}
It is easy to get the following specific heat and susceptibility
\begin{eqnarray}
C_{V}^{1f} &=&
- \frac{120}{\pi^3}\eta(\lambda)\,T^{2}\ln\frac{\Lambda}{T}, \\
\chi_{f}^{1f} &=&
- \frac{32}{\pi^3}\eta(\lambda)\,T\ln\frac{\Lambda}{T}.
\end{eqnarray}
Here, the ultraviolet cutoff $\Lambda$ can be taken to be of order
$10$eV, which is determined by $\sim a^{-1}$ with lattice constant
$a=2.46\textrm{\AA}$. From these results, we know that both specific
heat and susceptibility of massless Dirac fermions exhibit
logarithmic $T$-dependence due to long-range Coulomb interaction.
These are non-Fermi liquid behaviors.

The appearance of such singular fermion specific heat was first
pointed out by Vafek \cite{Vafek}. Here, we obtained the same
qualitative $T$-dependence by a different method. In Ref.
\cite{Vafek}, the calculation of free energy was performed on the
basis of the retarded vacuum polarization functions and retarded
fermion propagator $G^{{\rm ret}}(\omega,\mathbf{k}) =
\frac{1}{\omega-\sigma \cdot \mathbf{k}}$, while in our case the
polarization functions and fermion propagator are expressed in the
Matsubara formalism. Strictly speaking, these two polarization
functions are equivalent and should lead to the same results. We
numerically compute the free energy using both the polarization
functions obtained in the present paper and that in Ref.
\cite{Vafek}, and found that the results are very close to each
other (the maximum proportional error of the coefficient
$\frac{\delta f}{T^{3}\ln\frac{\Lambda}{T}}$ is $< 5\%$). In order
to get an analytic expression for free energy, some approximations
to the polarization functions is unavoidable. In Ref. \cite{Vafek},
the dominant contribution of polarization function comes from $y
\equiv \frac{\omega}{{\bf q}} \sim 1$ at both $y > 1$ and $y < 1$
regions (after analytic continuation the momentum becomes $q =
\sqrt{{\bf q}^{2} - (\omega + i\delta)^{2}} = {\bf q}\sqrt{1 -
y^{2}}$), while in our calculation the dominant momentum region is
$y \equiv \frac{\omega_{m}}{{\bf q}} \sim 0$ ($y \equiv
\frac{\omega}{{\bf q}} \sim 0$ in the continuous form). For this
reason, our analytic expression for the free energy differs from
that of Ref. \cite{Vafek} (the approximation of $J(x, y, q)$ might
partly explain the difference). After comparing the analytical
results with numerical results, we found that our analytical result
is slightly lower than the numerical result while the analytical
result in Ref. \cite{Vafek} is slightly greater than the numerical
result. For $\lambda = 1$, the analytical and numerical results for
the coefficient are $0.277$ and $0.231$ respectively in our work and
$0.200$ and $0.225$ respectively in Ref. \cite{Vafek}.

\begin{figure}[ht]
  \centering
    \includegraphics[width=2.7in]{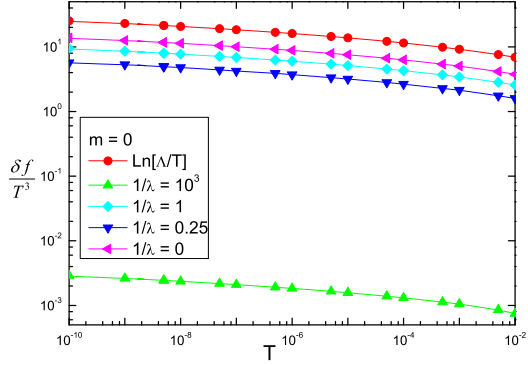}
\caption{Interaction correction to free energy in the semimetal
phase with $m=0$ for different interaction strength $\lambda$. The
red line is the reference free energy $\ln\frac{\Lambda}{T}$. It
appears that free energy displays the same logarithmic behavior for
different $\lambda$.} \label{fig:deltaf}
\end{figure}

The contribution of Coulomb interaction to the free energy in the
semimetal phase with $m=0$ is shown in Fig. \ref{fig:deltaf}. The
free energy behaves as $\propto T^{3}\ln T$ (logarithmic correction)
for several different values of $\lambda$.

\begin{figure}[ht]
  \centering
    \subfigure[]{
    \label{fig:spheat_m4}
    \includegraphics[width=2.7in]{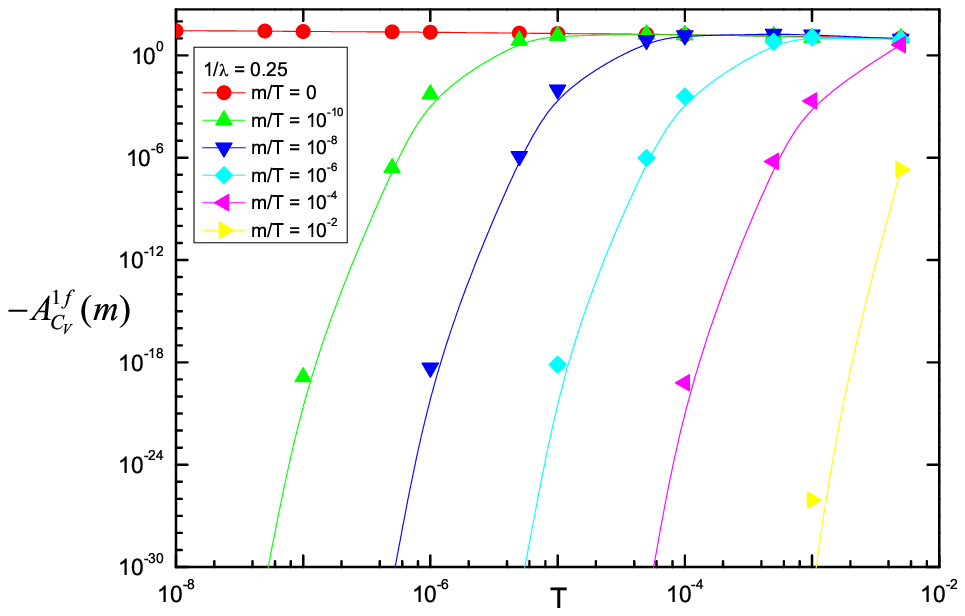}}
    \subfigure[]{
    \label{fig:sus_m4}
    \includegraphics[width=2.7in]{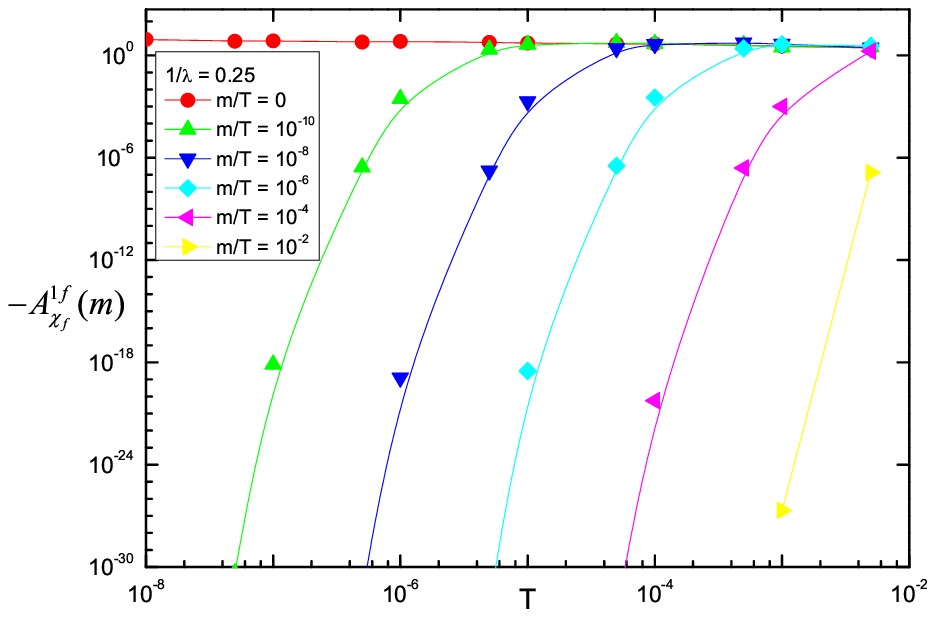}}
\caption{ (a) Interaction correction to fermion specific heat in the
insulator phase with $m \neq 0$ at $\lambda = 4$. (b) Interaction
correction to susceptibility in the insulator phase with $m \neq 0$
at $\lambda = 4$. The suppressing effect of fermion gap is
apparent.} \label{fig:tq_m}
\end{figure}

We now turn to the insulator phase where $m \neq 0$. The free energy
can be obtained by substituting Eq. (\ref{eq:pifull}) and Eq.
(\ref{eq:pi_0}) into Eq. (\ref{eq:delta_f}). The dependencies of
specific heat and susceptibility on different fermion mass $m$ for
$\lambda = 4$ are shown in Fig. \ref{fig:spheat_m4} and
\ref{fig:sus_m4}, respectively. The results for other choices of
$\lambda$ are similar and thus not shown. Here, we use the absolute
values $|\mathcal{A}^{1f}_{C_V}| = -\mathcal{A}^{1f}_{C_V}$ and
$|\mathcal{A}^{1f}_{\chi_{f}}| = -\mathcal{A}^{1f}_{\chi_{f}}$,
instead of $\mathcal{A}^{1f}_{C_V}$ and
$\mathcal{A}^{1f}_{\chi_{f}}$ which are negative. From Fig.
\ref{fig:spheat_m4} and \ref{fig:sus_m4}, we see that the fermion
gap leads to remarkable suppression of the interaction correction to
fermion specific heat and susceptibility.

In summary, in the semimetal phase the long-range Coulomb
interaction gives rise to non-Fermi liquid behavior of specific heat
and susceptibility. In the insulator phase, the fermion specific
heat and susceptibility are both significantly suppressed by the
excitonic gap, but the total specific heat has a finite value due to
the massless Goldstone bosons.

\section{Conclusion and discussion}
\label{sec:conclusion}

In this paper, we calculated the specific heat and susceptibility in
graphene. The ground state of graphene is semimetal when the Coulomb
interaction strength $\lambda < \lambda_c$, but becomes insulator
when $\lambda > \lambda_c$. The most prominent feature of semimetal
phase is the appearance of logarithmic $T$-dependence of specific
heat and susceptibility due to long-range Coulomb interaction. These
are non-Fermi liquid behaviors. In the insulating phase, because the
interaction correction to fermion excitations is strongly suppressed
by the excitonic gap, the total specific heat is solely determined
by the contribution from Goldstone bosons, while the susceptibility
drops significantly. Apparently, both specific heat and
susceptibility manifest quite different behaviors in the two sides
of the critical point $\lambda_c$.

Note that the semimetal and insulator phases both contain massless
excitations: massless Dirac fermion in the former and massless
Goldstone boson in the latter. They have different statistics and
exhibit completely different behaviors. For example, the massless
Dirac fermions can transfer heat current and produce a universal
thermal conductivity \cite{Durst} at $T = 0$, while the Goldstone
bosons only contribute a $\propto T^{3}$ term, which vanishes
rapidly as $T \rightarrow 0$. The massless Dirac fermions also gives
rise to a universal electric conductivity \cite{Durst}, although the
predicted electronic conductivity is at invariance with experimental
result (the famous missing $\pi$). The Goldstone bosons do not
contribute to electric conductivity since they are neutral.

We should point out that the Goldstone bosons are exactly massless
only when the Lagrangian respects a continuous chiral symmetry. If
the continuous chiral symmetry is explicitly broken by some contact
four-fermion interaction, then the Goldstone bosons are no longer
strictly massless. Instead, they have a small mass as the result of
dynamical breaking of appropriate continuous chiral symmetry
\cite{Weinberg}. In this case, our discussion and calculation about
the free energy contribution from Goldstone bosons should be
modified and a small mass should be included. In reality, there are
various four-fermion interactions in the graphene \cite{Alicea,
Gusynin09}. If the contact four-fermion interaction has the form
$\propto (\bar{\psi}\gamma_0 \psi)^2$, then the continuous chiral
symmetry is not explicitly broken and the Goldstone bosons are still
massless. If the four-fermion interaction term is $\propto
(\bar{\psi}\psi)^2$, then the system has only discrete chiral
symmetry and there are no massless Goldstone bosons \cite{Liu09,
Gusynin09}. Therefore, the specific heat of Goldstone bosons
presented in Sec.\ref{sec:clean} is valid only when the continuous
chiral symmetry is not explicitly broken by any four-fermion
interaction term.

We finally comment on the validity of $1/N$ expansion. The excitonic
insulating transition requires the Coulomb interaction between Dirac
fermions be sufficiently strong. In this strong coupling regime,
$1/N$ seems to be the only available expansion parameter, even if it
is not small ($N=2$ for graphene). In our specific case, the fermion
mass plays the dominant role in the insulator phase. It suppresses
significantly the Coulomb interaction contribution to fermion
specific heat. This implies that, within the $1/N$ expansion, the
next-to-leading order contribution could be neglected since it is
much less than the leading order contribution. It is reasonable to
speculate that higher order corrections in $1/N$ expansion are also
suppressed by the dynamical fermion mass. In the semimetal phase,
there is no such suppressing effect, so higher order corrections
might be important. As shown in the context, the analytical
calculation of next-to-leading order correction is already very
complicated, including higher order corrections will make analytical
calculation intractable. The specific heat of massless Dirac
fermions may be analyzed by renormalization group approach
\cite{Son07, Xu08}, which found power-law $T^{\beta}$ behavior after
summing up all orders of logarithmic corrections \cite{Son07, Xu08}.
However, the exponent $\beta$ can only be calculated by performing
$1/N$ expansion. Therefore, the validity of $1/N$ expansion also
needs to be studied in this approach.

\balance

\section{Acknowledgments}

We thank G. Cheng and J.-R. Wang for discussions. W. L. is grateful
to R. Asgari for very helpful correspondence. This work is supported
by National Science Foundation of China under Grant No. 10674122.


\begin{thebibliography}{99}
\bibitem{CastroNeto}
A. H. Castro Neto, F. Guinea, N. M. R. Peres, K. S. Novoselov, and
A. K. Geim, Rev. Mod. Phys. 81 (2009) 109 .

\bibitem{Gonzalez94}
J. Gonzalez, F. Guinea, and M. A. H. Vozmediano, Nucl. Phys. B 424 (1994) 595.

\bibitem{Khveshchenko2001}
D. V. Khveshchenko, Phys. Rev. Lett. 87 (2001) 246802; D. V.
Khveshchenko and H. Leal, Nucl. Phys. B 687 (2004) 323.

\bibitem{Gusynin02}
E. V. Gorbar, V. P. Gusynin, V. A. Miransky, and I. A. Shovkovy,
Phys. Rev. B 66 (2002) 045108.

\bibitem{Herbut06}
I. F. Herbut, Phys. Rev. Lett. 97 (2006) 146401.

\bibitem{Son07}
D. T. Son, Phys. Rev. B 75 (2007) 235423.

\bibitem{Vafek}
O. Vafek, Phys. Rev. Lett. 98 (2007) 216407.

\bibitem{Sheehy}
D. E. Sheehy and J. Schmalian, Phys. Rev. Lett. 99 (2007) 226803.

\bibitem{Gonzalez96}
J. Gonzalez, F. Guinea, and M. A. H. Vozmediano, Phys. Rev. Lett.
 77 (1996) 17.

\bibitem{Sarma2007}
S. Das Sarma, E. H. Hwang, and W. K. Tse, Phys. Rev. B 75 (2007)
121406.

\bibitem{Khveshchenko2006}
D. V. Khveshchenko and W. F. Shively, Phys. Rev. B 73 (2006) 115104;
D. V. Khveshchenko, J. Phys: Condens. Matter 21 (2009) 075303.

\bibitem{Hands08}
S. J. Hands and C. G. Strouthos, Phys. Rev. B 78 (2008) 165423.

\bibitem{Drut09}
J. E. Drut and T. A. Lahde, Phys. Rev. Lett. 102 (2009) 026802;
Phys. Rev. B 79 (2009) 165425.

\bibitem{Liu09}
G.-Z. Liu, W. Li, and G. Cheng, Phys. Rev. B 79 (2009) 205429.

\bibitem{Nambu}
Y. Nambu and G. Jona-Lasinio, Phys. Rev. 122 (1961) 345.

\bibitem{Appelquist86}
T. Appelquist, D. Nash, and L. C. R. Wijewardhana, Phys. Rev. D 33 (1986) 3704.

\bibitem{Appelquist88}
T. Appelquist, D. Nash, and L. C. R. Wijewardhana, Phys. Rev. Lett. 60 (1988) 2575.

\bibitem{Gusynin06}
V. P. Gusynin, S. G. Sharapov, and J. P. Carbotte, Phys. Rev. Lett.
96 (2006) 256802; V. P. Gusynin and S. G. Sharapov, Phys. Rev. B. 73
(2006) 254511; V. P. Gusynin, V. A. Miransky, S. G. Sharapov, and I.
A. Shovkovy, Phys. Rev. B. 74 (2006) 195429.

\bibitem{Kotov}
V. N. Kotov, B. Uchoa, and A. H. Castro Neto, Phys, Rev. B 80 (2009) 165424.

\bibitem{Asgari2009}
A. Qaiumzadeh and R. Asgari, New J. Phys. 11 (2009) 095023.

\bibitem{Liu09HTC}
H. Jiang, G.-Z. Liu, and G. Cheng, Phys. Rev. B 79 (2009) 174503.

\bibitem{Kaul}
R. K. Kaul and S. Sachdev, Phys. Rev. B 77 (2008) 155105.

\bibitem{Kapusta}
J. I. Kapusta and C. Gale, \emph{Finite-temperature field theory:
principles and applications}, (Cambridge, UK; New York,1994).

\bibitem{Asgari}
M. R. Ramezanali, M. M. Vazifeh, R. Asgari, M. Polini, and A. H.
MacDonald, J. Phys. A: Math. Theor. 42 (2009) 214015.

\bibitem{Chubukovprb2005}
A. V. Chubukov, D. L. Maslov, S. Gangadharaiah, and L. I. Glazman,
Phys. Rev. B 71 (2005) 205112.

\bibitem{Liuconfine}
G.-Z. Liu, W. Li, and G. Cheng, Nucl. Phys. B  825 (2010) 303.

\bibitem{Dorey}
N. Dorey and N. E. Mavromatos, Nucl. Phys. B 386 (1992) 614.

\bibitem{Durst}
P. A. Lee, Phys. Rev. Lett. {\bf 71}, 1887 (1993); A. Durst and P.
A. Lee, Phys. Rev. B {\bf 62}, 1270 (2000).

\bibitem{Weinberg}
S. Weinberg, \emph{The Quantum Theory of Fields}, Vol. II, Chap.19
(Cambridge University Press, 1996).

\bibitem{Alicea}
J. Alicea and M. P. A. Fisher, Phys. Rev. B 74 (2006) 075422.

\bibitem{Gusynin09}
O. V. Gamayun, E. V. Gorbar, and V. P. Gusynin, arXiv:0911.4878v1.

\bibitem{Xu08}
C. Xu, Y. Qi and S. Sachdev, Phys. Rev. B 78 (2008) 134507.


\end{thebibliography}
\end{document}